\documentclass[a4paper,12pt, reqno]{amsart}
\usepackage[utf8]{inputenc}
\usepackage{graphicx}
\usepackage{color}
\usepackage{array}
\newcolumntype{C}[1]{>{\centering\let\newline\\\arraybackslash\hspace{0pt}}m{#1}}

\usepackage[foot]{amsaddr}

\usepackage{subcaption}

\usepackage[usenames,dvipsnames]{xcolor}

\usepackage{tikz}
\usetikzlibrary{shapes,arrows,positioning}
\usetikzlibrary{decorations.markings}
\usetikzlibrary{decorations.pathreplacing}
\usetikzlibrary{decorations.pathmorphing,backgrounds,fit,petri}

\usepackage{geometry}
\usepackage{amsmath}
\usepackage{amsfonts}
\usepackage{amssymb}
\usepackage{amsthm}
\usepackage{cite}
\usepackage{float}
\usepackage{changepage}
\usepackage{nicefrac}

\author[]{Andreas I. Reppas$^1$, Georgios Lolas$^1$, Andreas Deutsch$^2$ and Haralampos Hatzikirou$^1$}
\address[]{$^1$ Center for Advancing Electronics Dresden, Technische Universit\"at Dresden, Germany \\
$^2$ Center for Information Services and High Performance Computing, Technische Universit\"at Dresden, Germany}
\title{The extrinsic noise effect on lateral inhibition differentiation waves}
\begin{document}
%\markboth{A. Reppas et al.}{The extrinsic noise effect on lateral inhibition differentiation waves}
%
%% Title portion
%\title{The extrinsic noise effect on lateral inhibition differentiation waves}
%\author{Andreas I. Reppas\footnote{These authors contributed equally to this work.}
%\affil{Center for Advancing Electronics, Technische Universit\"at Dresden, Germany}
%Georgios Lolas$^1$
%\affil{Center for Advancing Electronics, Technische Universit\"at Dresden, Germany}
%Andreas Deutsch
%\affil{Center for Information Services and High Performance Computing, Technische Universit\"at Dresden, Germany}
%and Haralampos Hatzikirou \footnote{Corresponding author: haralambos.hatzikirou@tu-dresden.de}
%\affil{Center for Advancing Electronics, Technische Universit\"at Dresden, Germany} }

\begin{abstract}
Multipotent differentiation, where cells adopt one of several cell fates, is a determinate and orchestrated procedure that often incorporates stochastic mechanisms in order to diversify cell types. How these stochastic phenomena interact to govern cell fate are poorly understood. Nonetheless, cell fate decision making procedure is mainly regulated through the activation of differentiation waves and associated signaling pathways. In the current work, we focus on the Notch/Delta signaling pathway which is not only known to trigger such waves but also is used to achieve the principle of lateral inhibition, i.e. a competition for exclusive fates through 
cross-signaling between neighboring cells. Such a process ensures unambiguous stochastic decisions influenced by intrinsic noise sources, e.g.~as ones found in the regulation of signaling pathways, 
and extrinsic stochastic fluctuations, attributed to micro-environmental factors. However, the 
effect of intrinsic and extrinsic noise on cell fate determination is an open problem. Our goal 
is to elucidate how the induction of extrinsic noise affects cell fate specification in a lateral inhibition mechanism. Using a stochastic Cellular Automaton with continuous state space, we show that 
extrinsic noise results in the emergence of steady-state furrow patterns of cells in a 
 {``}frustrated/transient" phenotypic state.
\end{abstract}

%\category{I.6.1}{Simulation and Modeling}{Simulation Theory}

%\terms{Design, Algorithms, Performance}

%\keywords{Cellular Automata, Differentiation Wave, Cell Fate, Notch/Delta, Noise}

%\acmformat{Andreas I. Reppas, Georgios Lolas, Andreas Deutsch, and Haralampos Hatzikirou, 2014.
%The extrinsic noise effect on lateral inhibition differentiation waves.}

%\begin{bottomstuff}
%The authors gratefully acknowledge support from the German Excellence Initiative via the Cluster of
%Excellence EXC 1056 Center for Advancing Electronics Dresden (cfAED).\\
%Haralampos Hatzikirou acknowledges the support of the BMBF grant eMED (SYSIMIT, grant 01ZX1308B).\\ 
%Authors' addresses: Andreas I. Reppas {and} Georgios Lolas
%{and} Haralampos Hatzikirou, Center for Advancing Electronics Dresden, Technische Universit\"at Dresden, Germany; Andreas Deutsch, Center for Information Services and High-Performance Computing, Technische Universit\"at Dresden, Germany
%\end{bottomstuff}

\maketitle

\section{Introduction}\label{Intro} 

Cell fate determination during developmental processes requires the integration of
lineage information and signaling cues at specific developmental time points to yield robust, 
reproducible cell fate executions. Most cell fate decisions are determinate 
and are typically coordinated by waves of differentiation. Such {``}waves of fate"  have been 
observed both during the development of the Drosophila visual system \cite{SatoSuz13} as well 
as in the vertebrate retina \cite{Cepko1996,Cepko2006}.
More specifically, for the Drosophila visual system differentiation waves are formed in different 
parts of the embryo following specific morphogenetic furrows in order to establish topographic 
connections throughout the fly visual system. Additionally, in retinal development, 
retinal progenitor cells undergo a series of state changes before they adopt a final differentiated 
fate while a state is defined by the cell competence to respond to both intrinsic and extrinsic cues.
The synchronization of these multiple differentiation waves is mediated by specific signaling pathways. 
One of the most important signaling pathways is the Notch/Delta \cite{Bray,Chitnis,Perron,Fortini,Lai,Cepko2006}.

The Notch/Delta pathway represents a juxtacrine signaling transduction mechanism in developmental biology 
for cell fate decisions, in particular in the nervous system \cite{TsakLouvi,TsakLake}. It is also known 
as the {``}lateral inhibition" or the {``}lateral specification" mechanism, where the selected cell blocks 
the ability of its neighbours to differentiate. More specifically, in lateral cell fate specification 
neighboring cells exchange signals in order to adopt a specific fate, while one {``}local" winner is selected. 
In this regard, in small domains a single {``}winner" is produced  whereas in larger fields a self-organizing 
{``}salt-pepper" or chessboard pattern is formed \cite{TsakLake}.

\begin{figure}[h]
\begin{center}
\includegraphics[width=0.6\linewidth, height=0.6\linewidth]{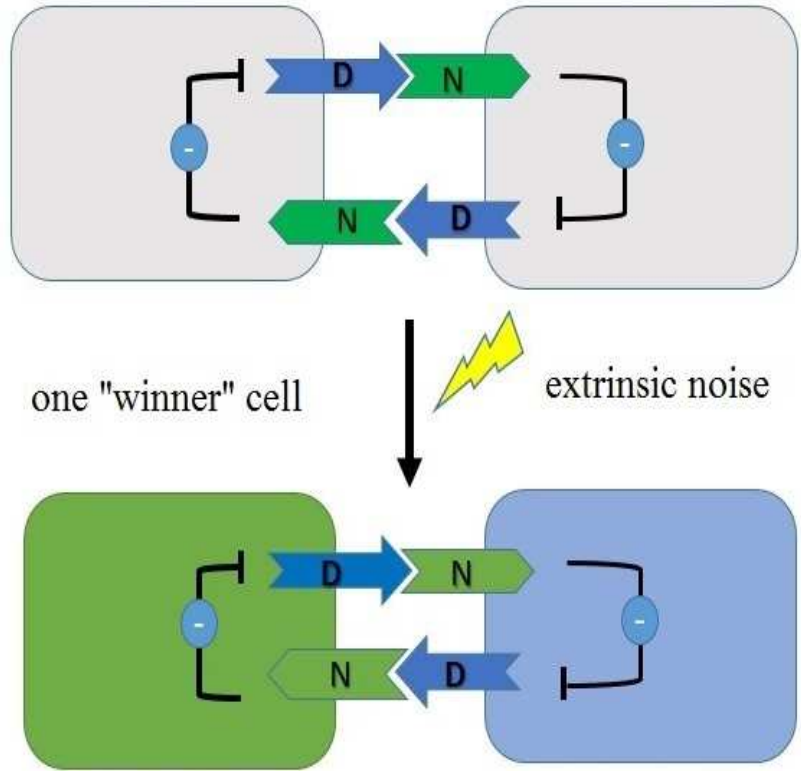}
\end{center}
\caption{\textbf{Notch/Delta mechanism}: The Notch receptor is activated upon binding to the Delta ligand 
that is anchored to the membrane of an adjacent cell.  The combination of intercellular Notch activation 
and intracellular Delta deactivation results in  two mutually exclusive fates of initially equipotent 
neighboring cells.}\label{fig:notchdelta}
\end{figure}

The main factors of the Notch/Delta pathway involve the extracellular domain of the two transmembrane 
ligands Delta and Serrate on the surface of one cell interacting with the extracellular domain of Notch 
receptor at the adjacent cell. The ligand/receptor binding triggers proteolytic events that result in
the cleavage of the receptor's intracellular domain. Notch interacts with Delta in two modes: activating
Delta signaling in neighboring cells (trans-activation) while inhibiting Delta signaling in the same cell
(cis-inhibition). Lateral inhibition patterning can emerge when Notch signaling downregulates Delta 
levels. Under specific conditions a high level of Notch in one cell will drive all of its neighbors to 
low levels of Notch and vise versa in the Delta case. This generates a stable lateral inhibition pattern
where each over-expressing Notch cell is surrounded by cells with increased Delta concentration 
(see Figure 1).

Although most of cell fate decisions are fixed within the developmental process, 
not all of them are determinate. Some cell fate decisions require random selection among alternatives. In agreement with determinate decisions, mechanisms acting after stochastic choices can compensate to yield robust outcomes \cite{losick2008,johnston2010}. Two stochastic influences can play crucial roles: the external and the 
internal noise. The intrinsic noise is associated with stochasticity involved in the genetic, epigenetic, transcriptional or translation regulations. On the other hand, extrinsic noise accounts for all the microenvironmental factors such as extracellular matrix or stromal components that can influence the cell fate determination process. In this work, we investigate the impact of extrinsic noise on the robustness of pattern formation  
introduced by the Notch/Delta mechanism. Here, the extrinsic noise coarse-grains the factors impacting Notch/Delta binding at the cells' membrane. We show that extrinsic noise can result in the 
formation of spatio-temporal {``}frustrated furrows" associating our results with recently discovered
biological evidence. Although the word noise is related with negative effects as something undesirable, 
in biological systems noise represents {``}randomness". In other words, noise may well be regarded as something 
desirable.

The structure of this paper is as follows: In section \ref{Models}, a thorough 
investigation of recently developed theoretical models of lateral inhibition is presented while in 
section \ref{MathModel} a simplified {``}lateral inhibition" model of Collier \textit{et al.} [1996]\nocite{Coll96} is defined. 
The patterning bifurcation diagram in a two-cell population is demonstrated in section \ref{Bifurc}. 
The fine grained {``}salt and pepper" pattern formation structure is presented and analyzed in subsection \ref{Numsimdeter}. 
A stochastic differential equation formulation for Notch-Delta is then derived in section \ref{stochcase}. In 
subsection \ref{numsimstoch}, our simulations results highlighting the emergence of a {``}frustrated" cell state 
associated with the effect of extrinsic noise are analyzed and compared with the deterministic case. 
In section \ref{biofrustrate} we investigate the preponderance of {``}frustrated/transient" cell fate mechanism in 
biological systems. A discussion together with suggestions for future work is finally given in 
section \ref{Conclude}.

\section{Modeling Lateral Specification}\label{Models} 

Several mathematical models have been developed to analyze lateral specification in different biological
contexts. The first model on juxtacrine signalling was formulated by Collier \textit{et al.} [1996]\nocite{Coll96} 
considering the role of a ligand, Delta, and its receptor, Notch, and demonstrating that lateral inhibition 
is able to generate fine-grained patterns. There have been other extended formulations of the aforementioned
model, by Owen and coworkers, either by considering the dynamics of ligands as well as free and bound 
receptors \cite{Webb04} or by inducing positive feedback realizations for both ligand and receptor 
\cite{WeaOwe,OweWea}, applied to different geometries (strings of cells and square and hexagonal arrays), 
and proving that lateral inhibition can generate patterns with longer wavelength. In a more recent paper, 
O{'}Dea and King [2011]\nocite{DeaKin} developed a multi-scale technique in order to construct a continuum 
model for investigating the pattern formation dynamics proposed in Collier \textit{et al.} [1996]\nocite{Coll96}.

Models dealing mostly with genetic circuits through the nonlinear lateral inhibition mechanism have
been proposed by Plahte [2001]\nocite{Plah01} and Shaya and Sprinzak [2011]\nocite{ShayaSprinzak}. 
More detailed models of the Notch/Delta circuit can be found in Hsu \textit{et al.} [2006]\nocite{Hsu06} and  
Meir at al. [2002]\nocite{Meir02}. More specifically, Hsu \textit{et al.} [2006]\nocite{Hsu06} demonstrated
that the lateral inhibition mechanism favors cells that are in contact with fewer inhibitory signal 
sending adjacent cells while Meir at al. [2002]\nocite{Meir02} investigated the parameter regime and 
how this affect the generation of new patterns through the lateral inhibition mechanism.

%Recently, Barad and co-workers proposed a probabilistic model in order to highlight the sources of 
%errors associated with the lateral inhibition process \cite{BarHor,BarRos}. They showed that lateral 
%inhibition is an error-prone process that suprisingly increases the fidelity of the process. 
%They argue that this beneficial effect of noise complements its role in the generation of phenotypic
%diversity. In other words, they revealed that the ability of Delta to inhibit Notch within the same g
%cell (the well known cis-interaction) reduces the error rate.

Just recently, several quantitative models have been developed in order to 
address specific experimental questions associated with the specific pathway \cite{Koizumi,Cohen_Baum,Cohen_Milan}. Koizumi \textit{et al.} [2012] studied the role 
of lateral inhibitory regulation on cells present at the leading edge (tip-cells) during Drosophila trachea development. 
Both their mathematical and numerical results revealed that Notch/Delta mechanism enhances 
the robustness of the tip-cell selection compared with a system regulated by self inhibition. 
The work by Cohen and coworkers \cite{Cohen_Baum,Cohen_Milan} showed that the development of the microchaete bristle pattern 
on the notum of the fruitfly Drosophila melanogaster is dependent upon a long-range interacting population mediated by dynamically extending actin-based filopodia. Cells use filopodia-like extensions to gather information from non-neighboring cells. Non-local cell interactions mediated by filopodia 
dynamics are thought to help cells receive morphogen signals as well as collect 
information regarding the identity of nearby cells that provide survival cues. 
These filopodia dynamics generate a type of structured noise that contributes to the formation of a well-ordered and spaced pattern of bristles by the induction of intermittent Notch-Delta signaling. 

An investigation of the role of structural noise (spatial as well as temporal) 
was presented in a more recent work of Cohen \textit{et al.} \cite{Cohen_Miodownik}. They proposed an asynchronous cellular automaton model to study a lateral inhibition mechanism in a hexagonal lattice. They showed that in the absence of noise they were able to recapitulate the results obtained from previous related continuous models using a 
discrete formulation without the need to invoke diffusible morphogens. Whereas the inclusion of noise 
can lead to stripped or spotted pattern refinement whose thickness is directly dependent on the noise signal threshold \cite{Cohen_Miodownik}.

Additionally, Sprinzak \textit{et al.} \cite{SprinElow} have studied Notch signaling from a synthetic biology perspective. 
In particular, they highlighted the issue of cis-inhibition effect, namely what happens if Notch and Delta 
bind in the same cell alongside with the trans-inhibition effect. In their seminal work, they developed 
different kinetic models that could reproduce the dynamic behavior of the engineered Notch-Delta circuits. 
Mainly, their models were reduced to three equations in which the dynamics between the ligand, the receptor 
and Notch intracellular domain are investigated.

It has recently been observed by Formosa and Ibanes \cite{Formosa2014} that cis-inhibition enriches 
cell patterning by promoting pattern multistability. More specifically, their work predicted  
novel characteristics of the Notch-Delta pathway such as the role of ligands that travel 
through the extracellular space and the role of diffusible morphogens on patterns wavefronts. Moreover, they showed the antagonistic role of Notch acting either 
as a lateral inducer or lateral inhibitor in specific developmental processes. For instance, 
they predicted that autoactivation of a specific proneural factor drives Delta upstream and
is a fundamental component for chicken inner ear robustness \cite{PetroFormosa2014} 

A recent work by Barad and co-workers highlighted that the ability of Delta to inhibit 
Notch within the same cell (cis-interaction) can reduce the sources of errors associated with the 
lateral inhibition process \cite{BarHor,BarRos}. They argue that this beneficial effect of 
noise complements its role in the generation of phenotypic diversity.

Several elements of the Notch/Delta pathway remain unclear. For example, the diffusible transport of ligands through the extracellular space suggest a new way of non-local cell-cell communication. Another example is the {``}bipolar" role of Notch/Delta. Namely, it is known that Notch can have both an inductive as well as an inhibitory role in specific organs. A thorough review of the aforementioned models as well as the latest scientific challenges in pattern formation through the Notch-Delta pathway can be found at \cite{Formosaphd}.

\subsection{Mathematical Model of Lateral Specification}\label{MathModel} 
  
In the current work, we construct a model based upon the mathematical framework proposed 
and analysed by Collier \textit{et al.} [1996]\nocite{Coll96}. The model is defined by a set of coupled ordinary 
differential equations that describe the dynamics of signaling mediated Notch-Delta intercellular activation 
and intracellular inhibition. A pair of ODE's governs the evolution of the levels of Notch and 
Delta activity in each cell \textit{i}  over time  $\tau$. The Collier \textit{et al.} [1996]\nocite{Coll96}
model in dimensionless form is given by:

\begin{subequations}
\begin{eqnarray}
\dot{n_{i}}=f\left(\bar{D}\right)-n_i \label{eq:nd1},\\
\dot{d_{i}}=g\left(n_i\right)-d_i,
\label{eq:nd2}
\end{eqnarray}
\end{subequations}
where $\bar{D}$ denotes the mean level of Delta activity in the cells surrounding cell \textit{i},
$\bar{D}=\frac{1}{|\mathfrak{N}_i|}\sum_{j\in \mathfrak{N}_i}{D_j}$, where $\mathfrak{N}_i$ denotes the neighbourhood
of cell \textit{i} and $|\mathfrak{N}_i|$ specifies the number of its neighbours. To model the lateral inhibition exerted
by one cell upon another, Collier \textit{et al.} [1996]\nocite{Coll96} defined the functions $f,g$ as:

\begin{subequations}
\begin{eqnarray}
f\left(x\right)=\frac{x^k}{\bar{\alpha}+x^k} \label{eq:nd3},\\
g\left(x\right)=\frac{1}{1+\beta x^h},
\label{eq:nd4}
\end{eqnarray}
\end{subequations}
where $f$ is a monotonically increasing function modeling the coupling between adjacent cells and $g$ is
a monotonically decreasing function representing the inhibitory effect of the bound Notch/Delta 
upon Delta production. The positive parameters $\bar{\alpha}, \beta, k, h$ determine the feedback strength. 
More specifically, the value of the exponent $k$ determines whether the Notch-Delta binding is monovalent 
($k \geq 1$) or cooperative ($k \geq 2$). Cooperative reaction means that the Notch receptor has more 
than one binding cites, so the receptor has the ability to bind to more than one ligand molecules while 
in the monovalent case ($k = 1$) the Notch receptor has only one binding site. Here, we assume the
monovalent case. In this regard, without loss of generality, we approximate function $f=a x$, noting that 
this latter formulation coincides with the Taylor expansion of equation (\ref{eq:nd3}) for $x \ll 1$ 
and $\alpha={\bar{\alpha}^{-1}}$. The reduced model is then given by:

\begin{subequations}
\begin{eqnarray}
\dot{n_{i}}=\alpha \bar{D}-n_i \label{eq:nd5},\\
\dot{d_{i}}=\frac{1}{1+\beta n_{i}^h}-d_i.
\label{eq:nd6}
\end{eqnarray}
\end{subequations}

%\emph{ We can nondimensionalize the system by setting $\tau=k_{1}t, n_i=N_i/N_0, d_i=D_i/D_0, \alpha=\Psi/{k_1}$
%and $u=k_1/k_2$. Assuming also that both Notch and Delta decay rates are similar speed we set $u=1$.
% The dimensionless equations can then be written in the follwoing form:}
%
Parameter $\alpha$ in equation (\ref{eq:nd5}) depicts the normalized strength of interaction between neighboring cells.
Since no explicit discussion of the biological motivation of the parameter estimation chosen by  Collier \textit{et al.} [1996]\nocite{Coll96}  was presented, we set the parameters:

\begin{equation}
h=5, \beta=100  \label{eq:nd7}
\end{equation}

By using a linear response of Notch production we seek to investigate the immediate effect 
of Delta concentration on the production of Notch activity (in the neighboring cells) by just tuning 
parameter $\alpha$ which served as our bifurcation parameter. 
Below, we will investigate the dynamics of pattern formation.

\section{Results}
In this section, we present the results of our analysis for a deterministic and a stochastic version of the model.

\subsection{Bifurcation analysis of the Deterministic case in the absence of External noise}\label{Bifurc} 

Firstly, we determine under which parameter regimes {``}salt and pepper" patterning is achieved with 
neighboring cells adopting opposing fates. Since the Notch/Delta mechanism under consideration is local,
analyzing the model equations for two cells with periodic boundary conditions gives insight into the 
period-two patterning behavior. Figure 2 presents the bifurcation diagrams of the Notch and Delta 
activation computed as fixed points of Equations \ref{eq:nd5},\ref{eq:nd6} with respect to the parameter 
$\alpha$. The bifurcation diagrams were extracted by performing an arc-length continuation method \cite{kelley1999}. The existence of a pitchfork bifurcation 
for a critical value of $\alpha$ leads to a bistable regime of Notch and Delta dynamics associated with a saddle point.

\begin{figure}[h!]
\begin{center}
\includegraphics[width=0.8\linewidth, height=0.4\linewidth]{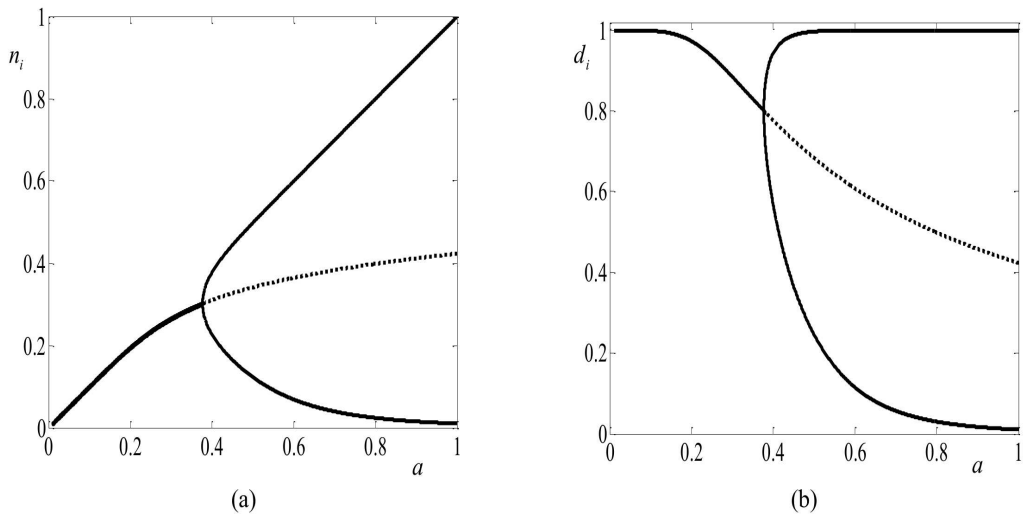}
\end{center}
\caption{\textbf{Bifurcation diagram}: Bifurcation diagram of the Notch concentration with 
respect to the strength of interaction, $\alpha$, in a pair of cells. 
The two stable solution curves (solid lines) in the central region correspond to the period-two 
fine-grained pattern, while the unstable state corresponds to the homogeneous state (dotted lines). 
The critical point which initiates  pattern formation is $\alpha_{cr}=0.3771$.}
\label{fig:bifurcation_initial}
\end{figure}

It can be seen, that if the strength of interaction is sufficiently small the only stable solution is the
homogeneous state where all the cells have the same concentration of Notch and Delta. At a critical
value, $\alpha_{cr}$, the homogeneous state loses stability and gives rise to a pair of heterogeneous steady
states where the cells adopt two distinct fates: high-Notch/low-Delta (primary fate) and low-Notch/high-Delta
(secondary fate). Thus, when $\alpha>\alpha_{cr}$ the two-cells system results in a mutual inactivation 
which creates the lateral inhibition pattern.

\subsubsection{Numerical simulations}
\label{Numsimdeter}
We consider a coupled map lattice model with continuous state defined on a two-dimensional regular lattice 
$L=L_1\times L_2$, where $L_1=L_2=\{1,...50\}$ are the lattice dimensions, 
with $2500$ nodes (cells) interacting in a von Neumann neighborhood with periodic boundary conditions. The state space of the model is $(n_i,d_i) \in [0,1]^2 \subset \Re^2$.
In order to study the fate specification wave we initialize our system by assuming a line of 
cells with opposite fates (inhibiting one another) while the rest of the cells are set to
a neutral state (no fate state), in other words having zero concentration of Notch and Delta 
(Figure 3(a)). The aforementioned model can be viewed as a 
Cellular Automaton with continuous state space coupled with synchronous update rules with the lateral inhibition process described by equations \ref{eq:nd1} and \ref{eq:nd2}. In other words, the transition rules associated with 
lateral inhibition can be represented by the following simple mechanism: the level of Notch 
activation, $n$, reflects the intensity of the inhibition the cell experiences while the 
level of Delta activity, $d$, reflects the intensity of the inhibitory signal that each cell delivers to its neighbors. More specifically, a cell surrounded by neighboring cells 
expressing high Notch is assumed to be in an inactive state (state 0), while a cell surrounded by neighboring 
cells expressing low Notch is assumed to be in an active state (state 1) In each case we performed simulation for 
parameter values $\alpha>\alpha_{cr}$ aiming to investigate how the fate specification 
wave propagates in space and time.

\begin{figure}[h!]
\begin{center}
\includegraphics[width=0.85\linewidth, height=0.6\linewidth]{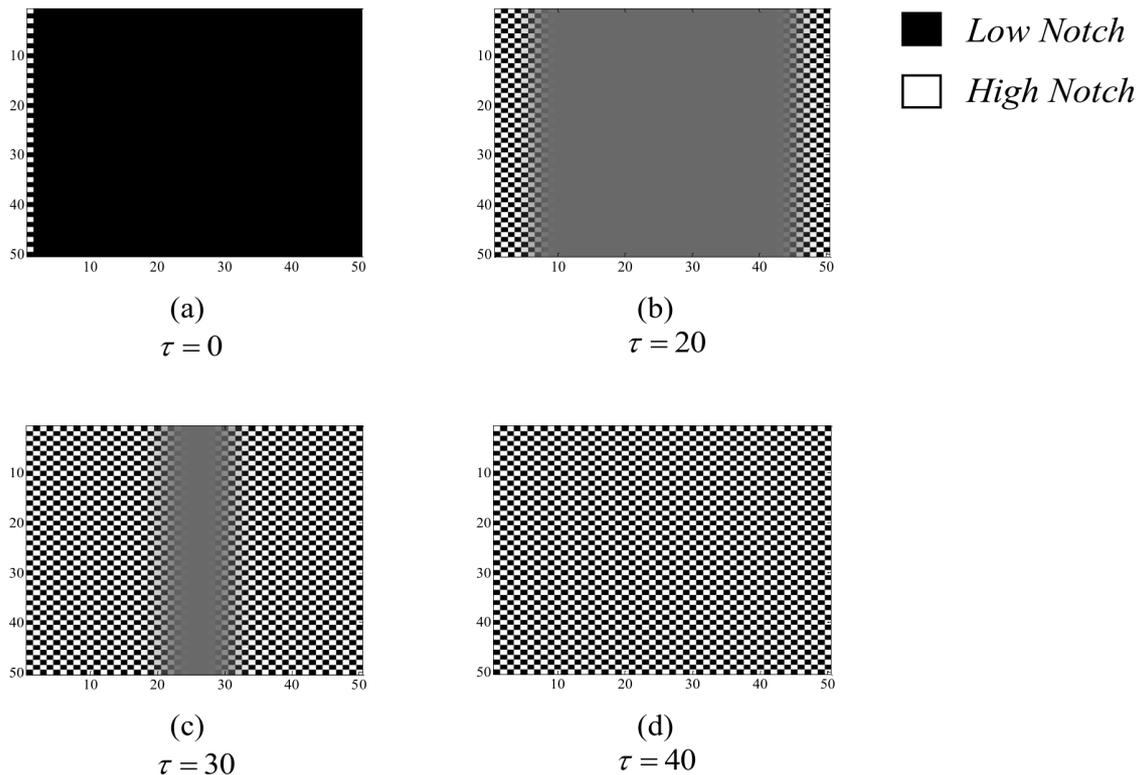}
\end{center}
\caption{\textbf{Deterministic fate specification wave:} The generation of a {``}fine-grained'' chessboard pattern in simulations of the deterministic model defined by the system of equations (\ref{eq:nd5}) and (\ref{eq:nd6}) with $\alpha=1$.}
\label{fig:deterministic_CA}
\end{figure}

We solve the system of coupled equations \ref{eq:nd5}, \ref{eq:nd6} numerically using a Runge-Kutta Method of $4^{th}$ order for each
of the $N=2500$ cells. Figure \ref{fig:deterministic_CA} presents the evolution of the fate wave specification when $\alpha=1$.
As expected, the initial line of fate specific cells evolves through time creating a fate wave that results in the
evolution of a {``}salt and pepper" pattern.

\begin{figure}[h!]
\begin{center}
\includegraphics[width=0.8\linewidth, height=0.45\linewidth]{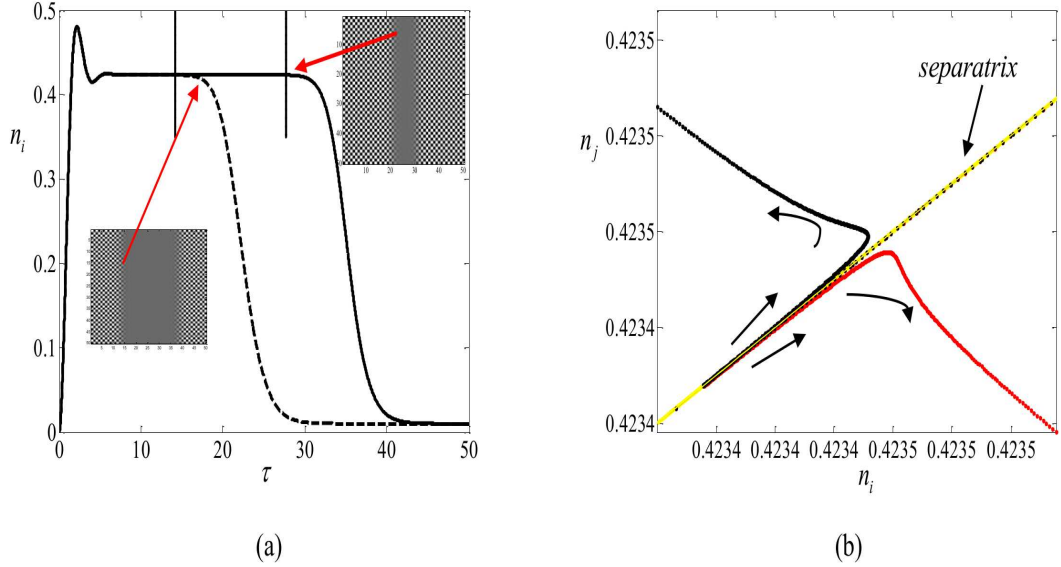}
\end{center}
\caption{\textbf{Impact of differentiation on cell dynamics:} (a) Time evolution of the Notch concentration of different cells. The vertical lines depict the time of fate selection for each cell. (b) Phase space of Notch concentrations between different cells. The trajectories correspond to consecutive neighboring couples of cells which adopt different fates. The yellow line is the stable manifold of the saddle node (see text for further information).}
\label{fig:deterministic_evolution_separatrix}
\end{figure}

In figure 4(a) we investigate the evolution of Notch concentration between two different cells. 
Spatial patterning is generated by a wave front, behind which a regular {``}salt and pepper" pattern forms. Initially, the level of Notch concentration in all cells (except the first line with cells having already 
decided their fate) is chosen close to the stable manifold of the saddle point (see Figure \ref{fig:bifurcation_initial}). Figure 4(b) depicts the phase space of the Notch concentration between two consecutive pairs of cells which finally adopt different fates. The stable manifold 
of the saddle point (denoted in yellow in the plane $(n_i,n_j)$ in Figure 4(b)) separates the attraction of the two different states \cite{Nayef2004}. The cells remain on the stable 
manifold until the wave fate specification forces them to follow the steady-state pattern.

\subsection{Stochastic case: The emergence of the third ``frustrated" state as a consequence of external noise}\label{stochcase}

Here, we investigate the impact of extrinsic noise on the pattern formation dynamics. 
We assume that extrinsic noise influences the cell's ability in receiving free Delta $\bar{D}$ and 
that the parameter $\alpha$ is a random variable $\alpha=\mu+\eta_{\tau}$, with $\mu\in[0,1]$ and 
$\eta_{\tau}\sim \mathcal{N}(0,\sigma^{2})$. The rest of the parameters are defined by equation 
(\ref{eq:nd7}) presented in section $2$. Thus, the stochastic version of the deterministic model 
defined in equations (\ref{eq:nd5}, \ref{eq:nd6}) is: 

\begin{subequations}
\begin{eqnarray}
\dot{n_{i}}=\mu\bar{D}-n_i+\bar{D} \eta_{\tau} \label{eq:nd8},\\
\dot{d_{i}}=\frac{1}{1+\beta n_{i}^h}-d_i
\label{eq:nd9}
\end{eqnarray}
\end{subequations}

\subsubsection{Numerical simulations}
\label{numsimstoch}
Following the simulation setting presented in section \ref{stochcase}, we simulate the evolution of the fate 
specification wave by initializing the system as in the deterministic case. For the numerical 
integration of equations (\ref{eq:nd8}) and (\ref{eq:nd9})  in each cell $i$ we used the 
Euler-Maruyama method with time step $d\tau=0.01$ \cite{maruyama1955}.

\begin{figure}[h!]
\begin{center}
\includegraphics[width=0.8\linewidth, height=0.6\linewidth]{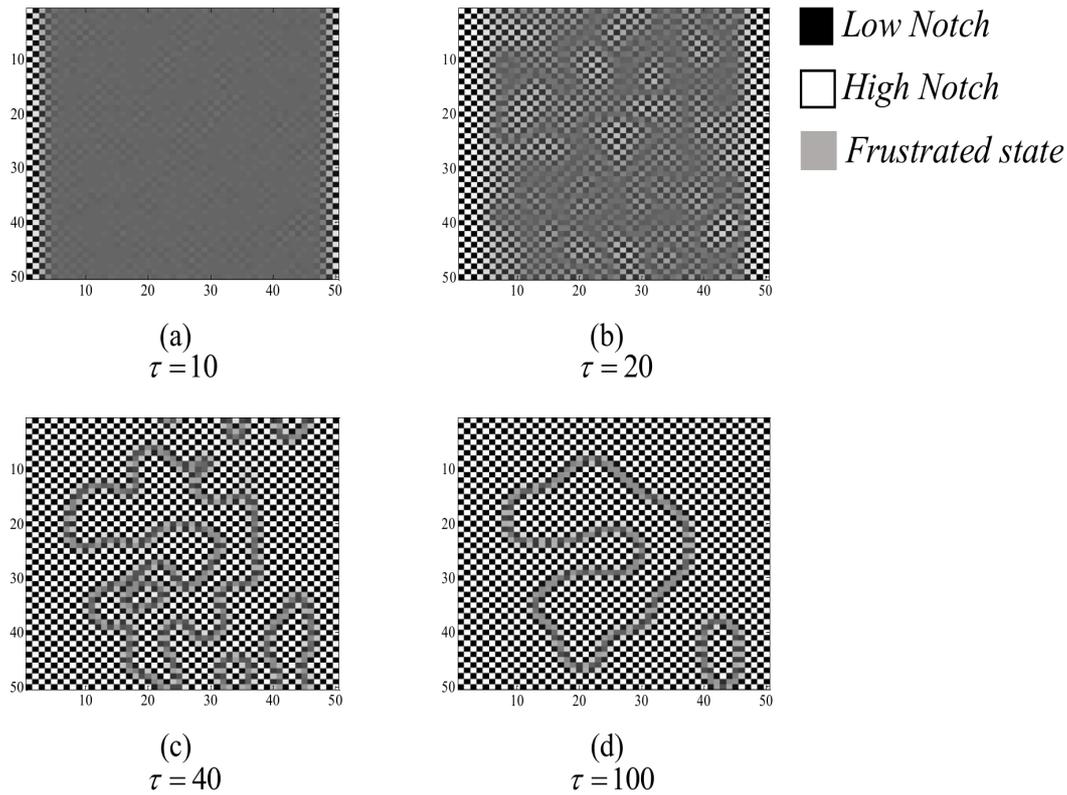}
\end{center}
\caption{\textbf{Emergence of frustrated states:} The evolution of  emerging {``}frustrated furrows'' 
within the {``}salt and pepper'' pattern in simulations of the stochastic  system of equations 
(\ref{eq:nd8}, \ref{eq:nd9}), with parameters $\mu=1$ and $\sigma=0.01$.}
\label{fig:noisy_CA}
\end{figure}

As expected, the extrinsic noise perturbs the steady state {``}salt and pepper" pattern found in the 
deterministic case. In this regard, in Figure \ref{fig:noisy_CA}, at $\tau=20$ we note that cells in the middle of 
the domain select a fate before the arrival of the differentiation wave. As time evolves in $\tau=40$, we observe that 
most of the cells have already been differentiated. However, several closed paths 
of {``}frustrated/transient cells", i.e.~undecided ones, emerge enclosing {``}chessboard" patterned cells. 
By time $\tau=100$, a combination of a fine grained pattern of {``}salt and pepper" and {``}furrows 
of frustrated cells" has emerged. 

\begin{figure}[h!]
\begin{center}
\includegraphics[width=0.8\linewidth, height=0.5\linewidth]{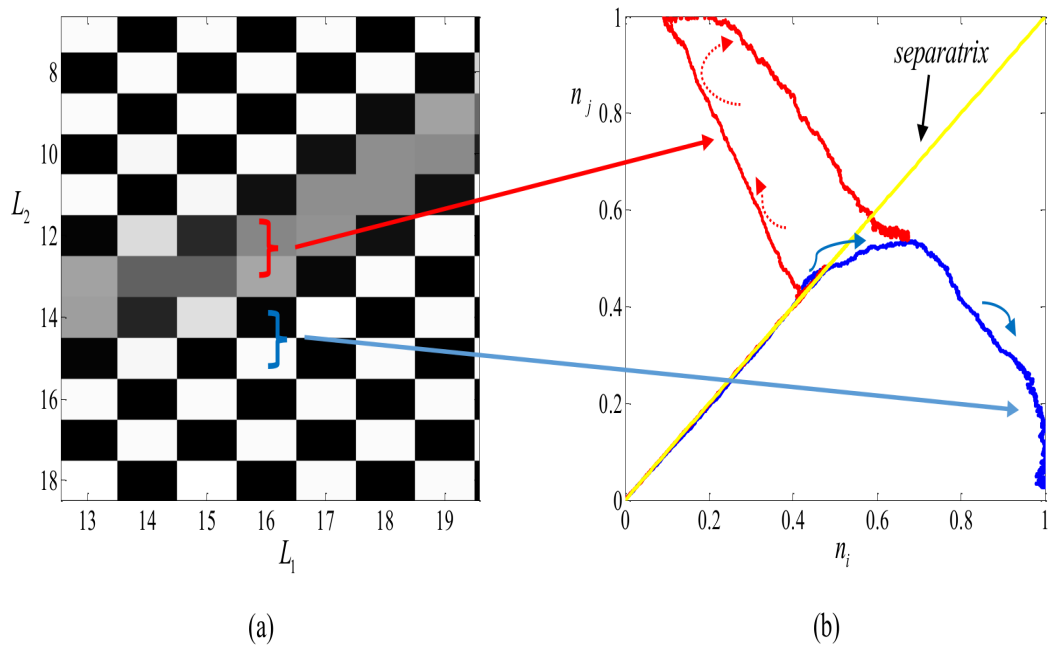}
\end{center}
\caption{\textbf{{``}Frustrated furrows'' dynamics:} (a) A close look to a neighborhood of a {``}frustrated 
furrow'' reveals cell fate switching. (b) Phase portrait of Notch concentrations between different couples 
of cells. See text for further details.}
\label{fig:frustrated_states_and_closer_view}
\end{figure}

By having a closer look at Figure 6(a), we see that these {``}frustrated furrows" are formed by cells 
that have adopted a fate which is different from the deterministic case. The exact formation of these 
furrows is varying for different system realizations. As we can observe in Figure 6(a), as soon as the 
differentiation wave has reached these frustrating furrows a conflict in cell fate determination arises: 
cells in the region within the ``frustrated furrows"  have opposite order of successive black and white 
nodes compared to cells in the region outside the furrows. Cells located at the vicinity of these two 
regions remain undecided, in terms of their fate specification, and therefore the frustrated/transient or mixed 
state perpetuates in time (frustrated cells denoted by gray in Figure 6(a)).

Additionally, in Figure 6(b) we highlight the Notch concentration of two neighboring cell pairs. In one 
pair (blue curve) noise forces the trajectory to cross the stable manifold, implying the selection of 
opposite cell fates compared to the deterministic case as shown in Figure 4(b) while the other pair of cells (red curve) stay close 
to the separatrix. Furthermore, in the stochastic model {``}fate selection"  takes place earlier than 
in the deterministic case (Figure 7). 

\begin{figure}[h!]
\begin{center}
\includegraphics[width=0.5\linewidth, height=0.4\linewidth]{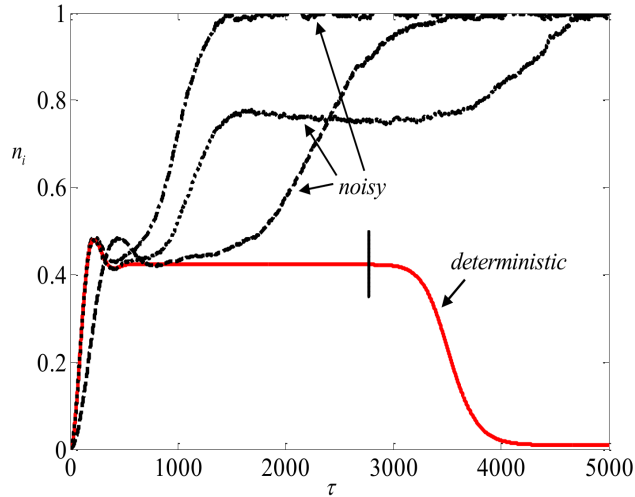}
\end{center}
\caption{\textbf{Effect of noise on cell fate specification:} Comparison of a cell's Notch level time evolution in  the presence of noise ($\mu=1$ and $\sigma=0.01$) and in  the deterministic case ($\alpha=\mu=1$ and $\sigma=0$). Stochastic fluctuations can induce cell fate selection before the arrival of the differentiation wave. The vertical line illustrates the time of cell fate decision in the deterministic case.}
\label{fig:differentiate_earlier}
\end{figure}

Finally, in Figure 8(a) and 8(c) we compare the state distribution
in the deterministic ($\mu=1$ and $\sigma=0$) and the stochastic case ($\mu=1$ and $\sigma=0.01$),
averaging over 100 simulations for time $\tau=100$. The intermediate states denote the existence of
 frustrated cells found in the {``}fate switching furrows". Additionally, in Figure 8(b) and 8(d), we compare the Notch dynamics in a
cross-section of the lattice for both cases. As expected,
in the deterministic case half of the cells adopt the high Notch fate while the rest adopt the low Notch one.
On the contrary, in the stochastic case, the emergence of frustrated cell furrows divides the domain into patterned regions.

We also studied the effect of varying the noise amplitude. A significant range of noise levels gave rise to emerging frustrated regions within 
the {``}salt and pepper" pattern. By increasing the level of noise beyond $\sigma>0.3$ we obtained a fully disordered system (Figure \ref{fig:disorder}). Additionally, we investigated the effect of neighborhood topology, i.e. the signaling communication range. For a Moore neighborhood (8 immediate neighbors) a refined stripped pattern is obtained   (Figure \ref{fig:neighbors8}(a)). In a similar manner, 
when noise is implemented frustrated furrows emerge within the stripped pattern, as shown in Figure \ref{fig:neighbors8}(b). The aforementioned results are in line with those obtained by Cohen \textit{et al.} [2010] using a hexagonal lattice. In the same study, the authors suggested that Notch/Delta spatiotemporal dynamics are similar to phenomena found in solid state physics, such as grain growth and recrystallization \cite{miodownik2001,miodownik2009}. Adding to this discussion point, our results suggest that dynamics of the frustrated patterns are similar with spin glasses phenomena. In the next section, we discuss the biological relevance of the Notch/Delta mechanism and the existence of frustrated/transient phenotypic cells states. 

\begin{figure}[ht!]
\begin{center}
\includegraphics[width=1\linewidth, height=1\linewidth]{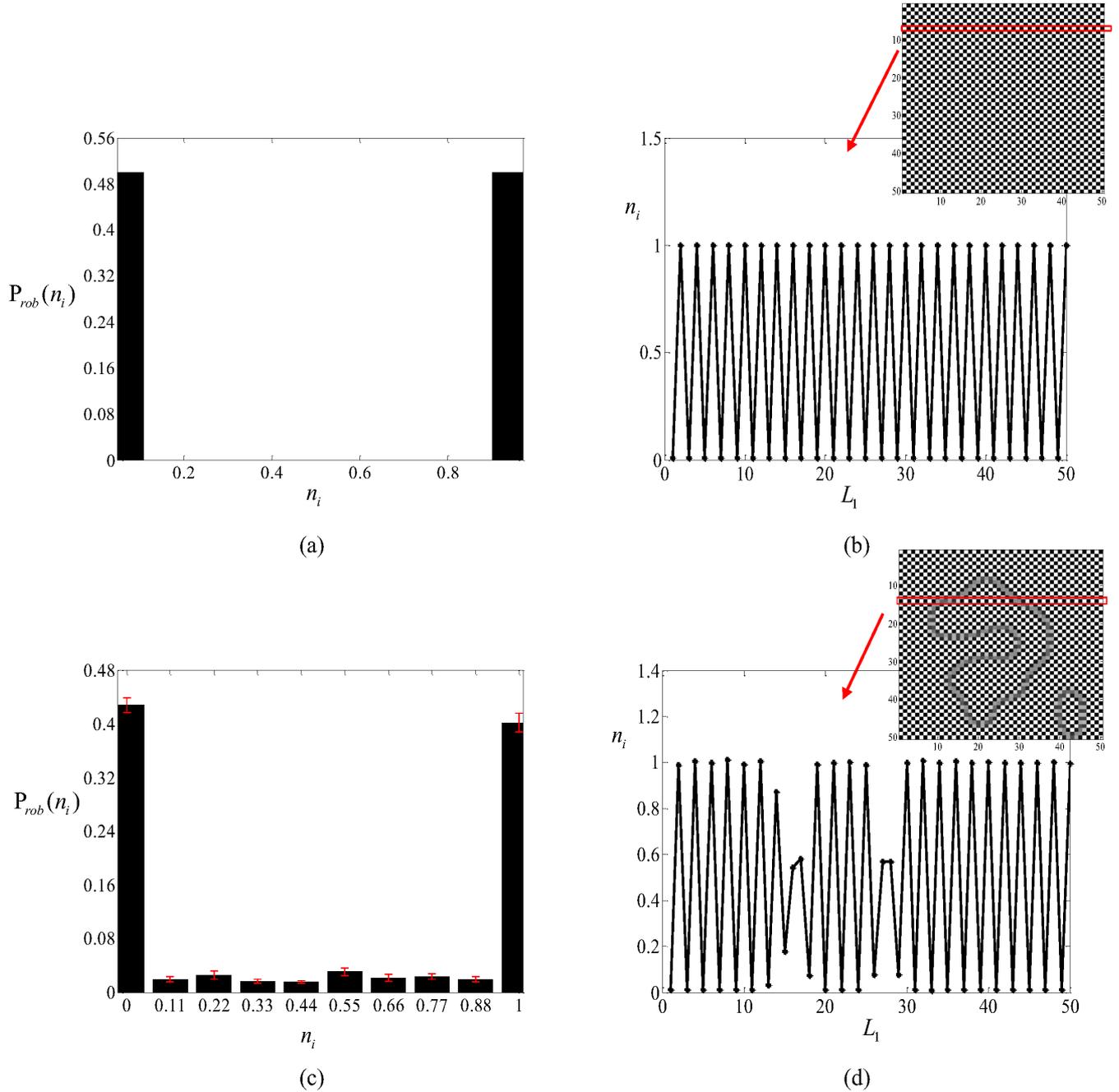}
\end{center}
\caption{\textbf{Notch probability distribution:} (a) Notch probability distribution in the 
deterministic case, when $\alpha=\mu=1$. Half of the cells adopt the high Notch fate and the 
other half the low one. (b) Notch level in one-dimensional cross sections for the deterministic 
case. (c) Notch probability distribution in the stochastic case for $\mu=1$ and $\sigma=0.01$ 
averaged over 100 simulations for time $\tau=100$. Some cells adopt an intermediate state. 
(d) Notch level in one-dimensional cross sections for the stochastic  case.}
\label{fig:distributions_correlations}
\end{figure}

\begin{figure}[h!]
\begin{center}
\includegraphics[width=0.5\linewidth, height=0.5\linewidth]{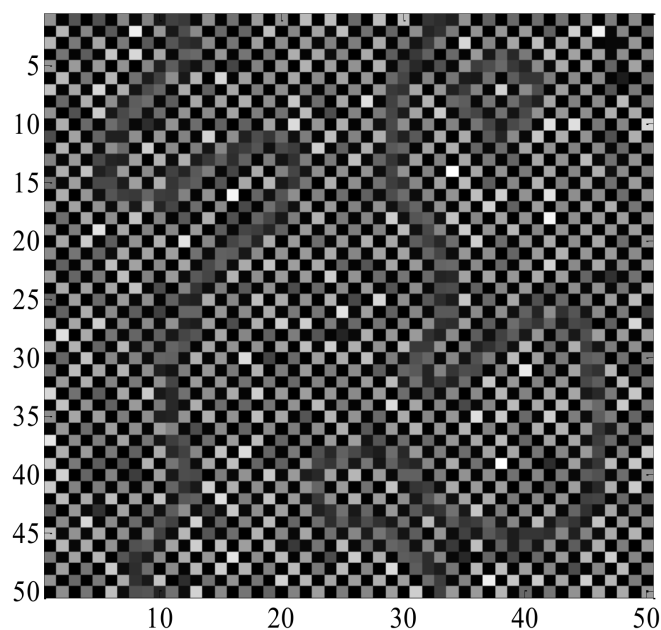}
\end{center}
\caption{\textbf{Disorder for high intensity noise:} Simulations of the stochastic system of equations 
(\ref{eq:nd8}, \ref{eq:nd9}), with parameters $\mu=1$ and $\sigma=0.35$. Furrows of cells in a {``}frustrated"/transient state evolve through the system resulting in a fully disordered pattern.}
\label{fig:disorder}
\end{figure}

\begin{figure}[ht!]
\begin{center}
\includegraphics[width=0.7\linewidth, height=0.4\linewidth]{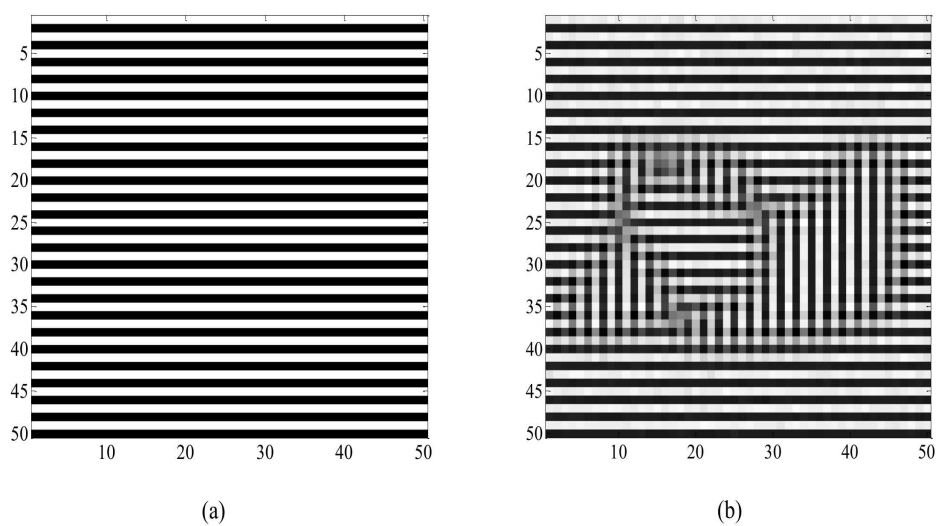}
\end{center}
\caption{\textbf{Neighborhood effect on the Notch/Delta model:}(a) A pattern of consecutive high Notch (low Delta) and low Notch (high Delta) stripes is obtained in the case we assume interactions with 8 neighbors for each cell. (b) The emergence of frustrated furrows in simulations of the stochastic system of equations 
(\ref{eq:nd8}, \ref{eq:nd9}), with parameters $\mu=1$ and $\sigma=0.02$ when each cell interacts with 8 neighbors.}
\label{fig:neighbors8}
\end{figure}

\subsection{The frustrated/mixed state in cell-fate decision making}
\label{biofrustrate}
On the basis of the above considerations, we have raised the possibility 
that during development cell decision making processes do not follow a deterministic 
chain of events in adopting their final fate. Instead, we propose that extrinsic noise in cell fate specification signaling may force cellular sub-population to a metastable, transient, frustrated state. Even though many of the components regarding differentiation 
programs are known, how noise may affect cell fate specification over time remains 
poorly understood. In this regard, it is not known if stochastic fluctuations are 
required before the decision point or after the decision has been made. 

Although a role for Notch/Delta signaling pathway as a lateral inhibitory mechanism is
well established, it has been a subject to considerable probing of many other of its attributes in retinal 
development. In this regard, apart from its well known role as a cell fate mechanism, Cepko and coworkers 
\cite{Cepko2006,Jadhav06} proved recently that it could also be recruited to allow a preservation of a pool of retinal progenitor cells(RPCs) in a transient/frustrated competent state for later cellular diversification. 
During this competent/transient state, cells interact with both inhibitory and stimulatory environmental 
(extrinsic) factors to reach a fate decision. In a similar manner, 
Shin \textit{et al.} \cite{Shin} investigated the role of Notch signaling in regulating a motoneuron-interneuron fate decision. They proposed that transient Notch inactivation disrupts lateral inhibition, 
committing more precursors to lineages that produce ventral spinal cords (also known as pMNs) and 
interneuron cells. Whereas continuous inhibition of Notch activity produces excess pMNs and a deficit of interneurons.

The notion of mixed cell fate decision echoes recent 
experimental studies of the cellular mechanisms underlying retinal epiphysis. Cau and 
coworkers \cite{CauBlad,CauQui} 
revealed a new functional role for the Notch-Delta pathway apart form its seminal 
role as the binary fate switcher. Namely, they proposed that Notch activation is required 
in order to seggregate epiphysial neurons that are in a transient phase of {``}double" or 
{``}mixed" identity. More specifically, in epiphysis there are two main type of cells; projection neurons 
and photoreceptors. However, a new type of cells has been identified of mixed 
identity. These cells have not yet received the postulated photoreceptor inducing signal while 
they have also not yet downregulated the projection neuron signal. Therefore, they remain in a mixed 
identity state. However, these mixed identity cells are rarely observed in wild-type epiphysis. Thus, it is speculated that either those cells do not represent a true cell {``}fate" or due to the lack of appropriate cell markers 
their {``}mixed" state is difficult to be distinguished.

Another example of a {``}mixed" cell phenotype is the intrinsically 
photosensitive retinal ganglion cells (ipRGCs) of the mammalian retina that reflect both the 
characteristics of projection neurons as well as photoreceptors \cite{BerDun,HatLia,DoKa,Aren}. 
ipRGCs are able to express the melanopsin photopigment while there are also photosensitive. 
In this regard, ipRGCs represent a potential example of a mixed/frustrated cell fate. It is 
speculated that retinal ganglion cells (RGCs) have lost through evolution their photosensitivity 
characteristic apart from their subfamily of ipRGCs that managed to retain this dual characteristic.  
It would be rather interesting to test biologically whether Notch activation can seggregate ipRGCs 
dual function. %However, it is not yet known if these examples reflect noise effects.

Our findings draw also parallels to a recent work by Alfonso Martinez Arias and coworkers \cite{MunozArias12,Munoz12}
that investigated the intermediate transition states that arise during cell fate 
decision processes as a result of the co-operation of two signaling pathways (Wnt and 
Notch). This transition state is an emerging highly unstable
and reversible state between the initial (pluripotent) and differentiated state.

Lately, Vistulo de Abreu and coworkers \cite{Abreu06,Abreu09} proposed that frustrated cells can 
provide another approach of functional immune system. They considered a system 
of three cells where they try to form stable conjugates. For example, if cells {``}A" and {``}B" are 
conjugated and cell {``}C" is alone, cell {``}C" can form a new conjugate with either cell {``}B" or {``}A", 
destroying the previous {``}AB" conjugate. Therefore, {``}A" becomes a frustrated cell by the presence 
of cell {``}C", and returns to the non-conjugate state. Every conjugate cell can be destabilized by 
interactions with neighboring cells. It will be rather interesting to investigate the Notch/Delta effect on the aforementioned 
immune system configuration. 

Finally, the existence of transient/frustrated cell phenotype has already been introduced 
in the epithelial to mesenchymal transition (EMT) \cite{Zhang}. During EMT, cells gain the ability to migrate 
and invade by loosing epithelial characteristics and gaining mesenchymal attributes. However, 
between the epithelial (E) and mesenchymal (M) states, there exists an intermediate 
phenotype known as the partial (P) EMT state. The P state retains both characteristics of
epithelial cells as well as features of mesenchymal states. Namely, the P state represents
a transient/frustrated phenotype that it could be transformed to E or M phenotype respectively. 
Recent experimental and theoretical models revealed that several cell types, starting 
from the epithelial phenotype, converted to partial EMT and then to mesenchymal phenotypes 
as concentration of transforming growth factor-$\beta$ (TGF-$\beta$) was increased.

The {``}frustrated states" identified in our model may also be associated with stochastic fate 
switching. The effect of noise in stochastic fate switching has already been 
observed and analyzed in the context of genetic circuits. More specifically, stabilizing and 
destabilizing roles of noise in genetic circuits have been reviewed in Eldar and 
Elowitz [2010]\nocite{EldarElo} (and references therein).    

The aforementioned models highlight that transient behavior can be regarded as an evolutionary novel feature of developmental growth. 

\section{Conclusions}
\label{Conclude}
The results and ideas discussed above suggest a general framework of the effect of 
extrinsic noise in the Notch/Delta induced cellular pattern formation. 
In particular, we have adopted a simplified version of the model described in Collier \textit{et al.} 
[1996]\nocite{Coll96}, which can produce a {``}salt and pepper" cellular pattern through the use of a Cellular Automaton formulation. We also investigated 
the impact of the differentiation wave, associated with the Notch/Delta mechanism, on the 
cell fate specification for the stochastic and deterministic case. In the deterministic case,
pattern formation is coordinated by a wave of differentiation. Whereas on the stochastic case extrinsic noise weakens the 
effect of the differentiation wave. Our analysis showed that even small noise intensities can give rise to {``}frustrated cell furrows".

However, the observation of ``frustrated states" seems to be more an exception than a rule in biological development.
Obviously, cells have developed de-noising mechanisms. An example is described in Barad et al. [2010]\nocite{BarRos} who identified the existence of error minimization mechanisms associated with the Notch/Delta pathway. Later, Sprinzak et al. [2010]\nocite{SprinElow}  proposed the \textit{cis-trans} Notch/Delta pathway which includes the exact ingredients for the suppression of stochastic fluctuations. Further design ideas for de-noising mechanisms can be expected from the analysis of extensions of the mathematical model introduced in this paper.

\bibliographystyle{amsedit}
\bibliography{bibliography_22_11}
%\lhead[]{Bibliography}

\end{document}